\documentclass{ws-ijmpd-mod}
\pdfoutput=1
\usepackage{bm}
\usepackage{amssymb}
\usepackage{pifont}
\usepackage{slashed} 
\usepackage{slantsc} 
\usepackage[usenames,dvipsnames,svgnames,table]{xcolor}

\def\d{{\mathrm{d}}}

\begin{document}
\title{
Quantum mechanix plus Newtonian gravity \\
violates the universality of free fall
}
\author{Matt Visser}
\address{School of Mathematics and Statistics\\
Victoria University of Wellington, PO Box 600, \\
Wellington 6140, New Zealand}
\maketitle
\begin{history}
\received{: 28 July 2017; 19 Sept 2017} 
\end{history}
\begin{abstract} 

\noindent
Classical point particles in Newtonian gravity obey, as they do in general relativity, the universality of free fall. 
However classical structured particles, (for instance with a mass quadrupole moment), need not obey the universality of free fall. 
Quantum mechanically, an elementary ``point'' particle {(in the particle physics sense)} can be described by a localized wave-packet, for which we can define a probability quadrupole moment.  This probability quadrupole can, under plausible hypotheses, affect the universality of free fall. {(So point-like elementary particles, in the particle physics sense, can and indeed must nevertheless have structure in the general relativistic sense once wave-packet effects are included.)} This raises an important issue of principle, as possible quantum violations of the universality of free fall would fundamentally impact on our ideas of what ``quantum gravity'' might look like. I will present an estimate of the size of the effect, and discuss where if at all it might be measured.

\bigskip
\noindent
{\sc Preprint}: 	arXiv:1705.05493 [gr-qc]

\bigskip
\noindent
{\sc Published}: International  Journal of  Modern Physics D26  (2017) 1743027

\bigskip
\noindent
{\sc doi}: https://doi.org/10.1142/S0218271817430271

\null\qquad\qquad\quad wave-packet; Newtonian gravity.

\vskip 10 pt
\noindent 
{\sc Date:} 10 March 2017; 15 May 2017; 28 July 2017; 20 September 2017; \\
\LaTeX-ed \today.

\vskip 10 pt
\noindent 
\leftline{Essay awarded an honourable mention in the 2017 Gravity Research Foundation Essay Competition.}

\end{abstract}

%
%

\bigskip
\hrule
\bigskip
\clearpage
\null
\vspace{-10pt}
\hrule
\markboth{Quantum mechanix plus Newtonian gravity violates the universality of free fall}{ }
\tableofcontents
\markboth{Quantum mechanix plus Newtonian gravity violates the universality of free fall}{ }
\medskip
\hrule
\markboth{Quantum mechanix plus Newtonian gravity violates the universality of free fall}{}
\def\d{{\mathrm{d}}}
\def\O{{\mathcal{O}}}
\def\omicron{o}
\def\CC{\mathbb{C}_0}
\def\N{\mathbb{N}}
\def\Z{\mathbb{Z}}
\def\Q{\mathbb{Q}}
\def\R{\mathbb{R}}
\def\C{\mathbb{C}}
\section{Introduction} 
\parindent0pt
\parskip5pt
\vspace{-10pt}
While classical point particles in Newtonian gravity obey the universality of free fall \cite{Su:1994,Schlippert:2014,Uzan:2010,Damour:2001,Damour:1995,Nordtvedt:1971}, the situation is more subtle for quantum wave-packets, for which  the finite size of the wave-packet implies the existence of a quantum probability quadrupole. (This in addition to any classical mass quadrupole; and dominant whenever the classical mass quadrupole can be arranged to cancel out in the physics.)\footnote{{Unfortunately ``point like'' to a particle physicist means something different than it does to a classical relativist. The particles of the standard model are ``point like'' down to at least $10^{-22}$ m,  but can very easily have wave-packets at the Angstrom scale, some 12 orders of magnitude larger. It is this absolutely unavoidable wave-packet contribution to the probability quadrupole, (and hence the mass quadrupole), that I will focus on in this essay.}}
Plausibly this quantum probability quadrupole will couple to gradients of tidal forces, (second derivatives of the local gravity, third derivatives of the Newtonian potential), in a well defined and calculable manner.\footnote{{See also a very recent attempt \cite{Seveso} at formulating a quantum weak equivalence principle in terms of the Fisher information matrix.}}$^{,}$\footnote{{See also an earlier analysis \cite{Viola,Onofrio} in terms of quantum states that do not have a classical limit.}} I will explore this possibility, and estimate the size of the effect.\footnote{{There are also potential effects due to spin. 
Certainly in the classical limit a spinning test particle will have its angular momentum couple to the spacetime metric, thereby leading (via the Mathisson--Papapetrou--Dixon equations) to deviations from geodesic motion. It is less than clear whether one should then apply these classical arguments to quantum spin.
}}

\section{Quantum structure of wave-packets} 

Suppose we have an elementary quantum particle that is described by some localized and normalizable wave-packet. 
Then we can take the probability density to satisfy $\int \rho(x) \; d^3 x = 1$.
Define the centre of probability, and the \emph{spread} of the wave-packet, by
\begin{equation}
\bar x^i = \int \rho(x) \; x^i \; d^3 x;  \qquad \sigma^2 = \int \rho(x) \; (x^i-\bar x^i)^2 \; d^3 x.
\end{equation} 
Now define the dimensionless probability quadrupole moment by
\begin{equation}
Q^{ij} = {1\over \sigma^2} \; \int \rho(x) \; (x^i-\bar x^i)(x^j-\bar x^j) \; d^3 x.
\end{equation} 
The trace of this probability quadrupole is automatically unity; implying that a traceless probability quadrupole can be defined by $\overline Q^{ij} = Q^{ij} - {1\over3} \delta^{ij}$. 
(Higher-order  multi-pole moments can be added as desired.)

\section{Gravitational force on a wave-packet}

For a classical point particle located at position $x^i$ the Newtonian gravitational force is simply $F_i = -m \nabla_i \phi(x)$.
For analyzing a quantum wave-packet we will have to make some assumptions. 

\vspace{-10pt}
One very natural assumption is that the \emph{net} force is simply given by integrating over the wave-packet weighted by the normalized probability density\footnote{{This is effectively an appeal to a minor variant of the Ehrenfest theorem, in the form $\langle F(x) \rangle =  \langle - \nabla V(x) \rangle$.
The analysis herein can be viewed as a refinement of the Ehrenfest theorem, wherein we construct an effective potential such that $ \langle \nabla V(x)\rangle =   \nabla V_\mathrm{effective}(\bar x,\sigma,Q)$, with the effective potential dependent on the centre, spread, and shape of the wave-packet.}}
\begin{equation}
(F_{net})_i = - m  \int \rho(x) \; \nabla_i \phi(x) \; d^3x.
\end{equation}
This sort of assumption is very much in line with the quite standard ideas espoused in setting up the Schr\"odinger--Newton equation~\cite{schrodinger-newton}.  To avoid this sort of result, one could for instance adopt a variant on Roger Penrose's ideas of a gravity-induced collapse of the wave function~\cite{emperor,shadows,reality}, or a Di\"osi-like approach~\cite{Diosi:1984,Diosi:1987,Diosi:2014}, or adopt a GRW variant~\cite{GRW}, or possibly some variant of a Bohmian approach~\cite{bohm1,bohm2}. In such a case $(F_{net})_i \to \langle F_{net}\rangle_i$ would presumably become an ensemble average over experimental outcomes. Be that as it may, for now I shall stay with the more-or-less standard interpretation of $(F_{net})_i $ as a \emph{net} force on an uncollapsed wave-packet, and see where that leads.

Now Taylor-series  expand around the centre of probability. We see
\begin{eqnarray}
(F_{net})_i &=& -m  \int \rho(x) \Bigg\{  \nabla_i \phi(\bar x)  + \nabla_i\nabla_j \phi(\bar x)\; [x^i-\bar x^i] 
\nonumber\\
&&\qquad\qquad
+ {1\over2}  \nabla_i\nabla_j \nabla_k \phi(\bar x)\; [x^j-\bar x^j] [x^k-\bar x^k] + \dots  \Bigg\} \; d^3x.
\end{eqnarray}
The integrals are easy, the probability dipole contribution vanishing in the usual manner:
\begin{equation}
(F_{net})_i = -m  \left\{  \nabla_i \phi(\bar x) 
+ {\sigma^2\; Q^{jk}\over2} \; \nabla_i\nabla_j \nabla_k \phi(\bar x)  + \dots  \right\}.
\end{equation}
The \emph{net} acceleration is now:
\begin{equation}
(a_{net})_i =  -\left\{  \nabla_i \phi(\bar x) 
+ {\sigma^2  \; Q^{jk} \over2}  \;\nabla_i\nabla_j \nabla_k \phi(\bar x)+ \dots  \right\}. 
\end{equation}
The bad news is that this is not universal --- \emph{net} acceleration depends both on the \emph{spread} $\sigma$ and the \emph{shape} $Q^{jk}$ of the wave-packet.
Even for two wave-packets with the same centre of probability,  acceleration differences will depend on differences in spread and shape. 
\begin{equation}
\Delta(a_{net})_i =  -\left\{  
 {\Delta(\sigma^2  \; Q^{jk}) \over2} \; \nabla_i\nabla_j \nabla_k \phi(\bar x)+ \dots  \right\}. 
\end{equation}

One simplification is to split the quadrupole into trace and trace-free parts, so that
\begin{equation}
(a_{net})_i =  \left\{  \nabla_i \phi(\bar x) 
+ {\sigma^2 \; \overline Q^{jk} \over2} \; \nabla_i\nabla_j \nabla_k \phi(\bar x) +
{\sigma^2\over6} \;  \nabla_i\nabla^2  \phi(\bar x) + \dots  \right\}.
\end{equation}
In empty space Laplace's equation $\nabla^2\phi=0$ implies that the last term drops out and
\begin{equation}
(a_{net})_i =  \left\{  \nabla_i \phi(\bar x) 
+ {\sigma^2 \; \overline Q^{jk} \over2 } \;  \nabla_i\nabla_j \nabla_k \phi(\bar x) \;+ \dots  \right\} 
\end{equation}
So we need to estimate both the spread of the wave-packet and the trace-free part of the probability quadrupole.
Note in particular that for any spherically symmetric wave-packet we have $\overline Q^{jk} =0$; then the quadrupole vanishes and the effect goes away. 
Furthermore, observe that this effect is not any usual notion of tidal effect --- the usual tides are governed by $\nabla_i\nabla_j\phi(\bar x)$ and are 2nd-order in gradients, and would have to do with internal stresses on the wave-packet --- this effect is 3rd-order in the Newtonian potential gradient.

\vspace{-10pt}
\section{Classical point source with quantum probe}
\vspace{-10pt}
Consider now a classical point source for the externally imposed Newtonian potential $\phi = GM/r$. Then
\begin{equation}
\phi \propto {1\over r};  \qquad \nabla_i \phi \propto {r_i\over r^3}; \qquad 
\nabla_i \nabla_j \phi \propto {\delta_{ij}r^2-3r_ir_j\over r^5};
\end{equation}
and finally
\begin{equation}
\nabla_i\nabla_j\nabla_k \phi \propto -{3 (\delta_{ij} r_k+\delta_{ik} r_j + \delta_{jk} r_i)\over r^5} +{15 r_i r_j r_k \over r^7}.
\end{equation}
Working in terms of unit vectors this becomes
\begin{equation}
\nabla_i\nabla_j\nabla_k \phi \propto -{3 (\delta_{ij} \hat r_k+\delta_{ik} \hat r_j + \delta_{jk} \hat r_i)\over r^4} +
{15 \hat r_i \hat r_j \hat r_k \over r^4}.
\end{equation}

So if the gravitational field is generated by a classical point source then the acceleration of the wave-packet is
\begin{equation}
(a_{net})_i =   -{GM\over r^2} \left\{ \hat r_i + {\sigma^2\; \overline Q^{jk}\over 2 r^2} 
[-3 (\delta_{ij} \hat r_k+\delta_{ik} \hat r_j + \delta_{jk} \hat r_i) +15 \hat r_i \hat r_j \hat r_k ] +\dots \right\},\;\;
\end{equation}
which  simplifies considerably
\begin{equation}
(a_{net})_i =   -{GM\over r^2} \left\{ \hat r_i 
\left[ 1+ {15\over2} \,{\sigma^2\over r^2} \,\{ \overline Q^{jk}\hat r_j \hat r_k \}\right]
- 3\;{\sigma^2\over r^2} \;\overline Q_i{}^{k} \hat r_k  +\dots \right\}.
\end{equation}

So the extra terms appearing in the net acceleration are of relative order $\mathcal{O}(\sigma^2/r^2)$, multiplied by dimensionless factors of order unity, and unless one of the principal axes of the wave-packet is aligned with the vertical, do not necessarily seem to represent a ``central force''. The ``sideways'' acceleration is perhaps a little less mysterious if one rewrites the acceleration as
\begin{equation}
(a_{net})_i =   GM \; \nabla_i \left\{ {1\over r}
\left[ 1+ {3\over2} \,{\sigma^2\over r^2} \,\{ \overline Q^{jk}\hat r_j \hat r_k \}\right]
  +\dots \right\}.
\end{equation}
\enlargethispage{20pt}
This is valid as long as the wave-packet is not appreciably evolving on the timescale $\tau$ set by the experiment, 
\begin{equation}
\tau \;\dot{\overline Q}^{jk} \ll {\overline Q}^{jk}.
\end{equation}
The net force is still a conservative potential force, but now  with an ``effective potential energy''
\begin{equation}
V_\mathrm{effective}  =  - {GMm\over r}
\left[ 1+ {3\over2} \,{\sigma^2\over r^2} \,\{ \overline Q^{jk}\hat r_j \hat r_k \}\right]
  +\dots.
\end{equation}

\section{Quantum source and quantum probe}

If both the source and the probe are described by quantum wave-packets then both source and probe have independent centres of probability, and independent probability quadrupoles. Then, taking $r$ to be the distance between the centres of probability, including the first quantum correction to the classical Newtonian potential implies
\begin{equation}
V_\mathrm{effective}  =  - {GMm\over r}
\left[ 1+ {3\over2} \;{
(\sigma_\mathrm{source}^2 \; \overline Q_\mathrm{source}^{jk}
+\sigma_\mathrm{probe}^2 \; \overline Q_\mathrm{probe}^{jk}) \;  \hat r_j \hat r_k
\over r^2} \right]
  +\dots.
\end{equation}
So the analysis is symmetric under the interchange of source and probe. 
For the quantum $N$-body problem, let $m_a$ be the mass, $Q_a^{jk}$ the probability quadrupole, and $(r_{ab})_i$ the relative displacements between the centres of probability of the individual wave-packets. Then to leading order
\begin{equation}
V_\mathrm{effective}  =  -\sum_{a\neq b} {Gm_am_b\over 2\, r_{ab}}
\left[ 1+ {3\over2} \;{
(\sigma_a^2 \; \overline Q_a^{jk}
+\sigma_b^2 \; \overline Q_b^{jk}) \;  \widehat{(r_{ab})}_j  \; \widehat{(r_{ab})}_k
\over r_{ab}^2} \right]
  +\dots.
\end{equation}
This is nicely symmetric under the interchange of any two of the wave-packets.\footnote{{While the Pauli exclusion principle implies that the wavefunction for fermions is odd under particle (wave-packet) interchange $\psi(x_a,x_b)=- \psi(x_b,x_a)$ the probability density $\rho(x_a,x_b) = |\psi(x_a,x_b)|^2=|\psi(x_b,x_a)|^2= \rho(x_b,x_a)$ is even.  So the effective potential is always even under wave-packet interchange.}}

\section{Experimental estimates} 
Let us now consider some rough estimates regarding the experimental/observational situation:
\begin{itemize}
\item 
For a nano-scale wave-packet ($\sigma \sim 10^{-9} \;m$) in the gravitational field of the Earth, (an idealized point Earth, $r\sim 6.371 \times 10^6 \;m$), we have
\begin{equation}
{\sigma\over r} \sim 10^{-16}; \qquad \left({\sigma\over r}\right)^2 \sim 10^{-32}.
\end{equation}
Now present-day E\"otv\"os-type experiments are extremely good~\cite{Dent:2008,Haugan:1976,Haugan:1977,Kraiselburd:2011}, (with the most precise experiments reporting a sensitivity of order $10^{-14}$), but they are still nowhere near 
good enough to have any hope of seeing this wave-packet effect.
\item 
For a typical laboratory-scale Cavendish experiment, $r \sim 10 \; cm = 10^{-1}\; m$, so for a nano-scale wave-packet  ($\sigma \sim 10^{-9} \;m$) we have
\begin{equation}
{\sigma\over r} \sim 10^{-8}; \qquad \left({\sigma\over r}\right)^2 \sim 10^{-16}.
\end{equation}
Detecting effects due to wave-packet structure still looks rather hopeless.

\item 
Proposed meso-scale Cavendish experiments~\cite{Schmole:2016}, are aiming for sub-millimetre distance-scales,   $r \lesssim 1 \; mm = 10^{-3}\; m$, so for a nano-scale wave-packet  ($\sigma \sim 10^{-9} \;m$) we have
\begin{equation}
{\sigma\over r} \sim 10^{-6}; \qquad \left({\sigma\over r}\right)^2 \sim 10^{-12}.
\end{equation}
For detecting a probability quadrupole, this still looks rather difficult.

\item
Only for a fully quantum Cavendish experiment,  might one eventually hope to get wave-packet separations of order the wave-packet spread. In that situation the effects due to wave-packet structure  would be of order unity; $\mathcal{O}(1)$. 
(So that the effects explored in this essay simply could not be ignored.) Achieving such sensitivity would be a very challenging experimental proposal. 
\end{itemize}
The current proposal, looking for effects of the probability quadrupole associated with a wave-packet, is orthogonal to currently extant experiments:
\begin{itemize}
\item 
The COW experiment~\cite{COW1,COW2}, and its variants, look at quantum interference of a split particle-beam on scales of $\sim 1 m$; but one is not dealing with wave-packets \emph{per~se}, and the COW experiments are insensitive to the probability quadrupole.
\item 
The semi-classical Cavendish experiment reported by Rosi \emph{et~al}~\cite{Rosi-et-al}, is based on a gradiometer measuring acceleration differences between clumps of $Rb$ atoms that are macroscopically separated on a scale~$\sim 33 \; cm$. (The shape of the $Rb$ clumps is unspecified, but if they are spherical, the entire quadrupole effect quietly vanishes.)
\item 
Bouncing neutrons off the floor~\cite{bounce1,bounce2,bounce3} is logically orthogonal to the quantum quadrupole effect for a different reason: There one is interested in probing the wave-function (built out of linear combinations of Airy functions) directly; looking for quantization effects that do not depend on a multipole expansion.
\item
The Page--Geilker experiment~\cite{page} is essentially a probe of the ``collapse of the wave-function",
and strongly suggests that semi-classical quantum gravity is not the whole story, (at least when working on a macroscopic scale of some several metres)~\cite{Kiefer:2008}. 
\end{itemize}

\section{Discussion} 

While certainly challenging, experimentally probing the ideas developed in this essay is not entirely impossible or implausible. Better, it is almost a no lose proposition --- if the effect is looked for and seen, (with an experiment of suitable sensitivity), then the quantum violations of the universality of free fall are certainly telling us something fundamental concerning the gravity-quantum interface. Conversely if the effect is found to not be there, it is most likely telling us that  gravity collapses the wave-function, \emph{a~la} Penrose--Di\"osi--GRW, since then integrating over the wave-packet to find the net force is not the appropriate thing to do. 
Either way, this would be a major step forward.

\bigskip
\centerline{---\,\#\#\#\,---}

\section*{Acknowledgments}
I wish to thank Eric Poisson for useful comments.

This research was supported by the Marsden Fund, through a grant administered by the Royal Society of New Zealand.



\begin{thebibliography}{69}

\bibitem{Su:1994}
  Y.~Su, B.~R.~Heckel, E.~G.~Adelberger, J.~H.~Gundlach, M.~Harris, G.~L.~Smith, and \hfill\\ H.~E.~Swanson,
  ``New tests of the universality of free fall'',\\
  Phys.\ Rev.\ D {\bf 50} (1994) 3614.
  doi:10.1103/PhysRevD.50.3614
  
  \bibitem{Schlippert:2014}
  D.~Schlippert {\it et al.},
  ``Quantum test of the universality of free fall'',\\
  Phys.\ Rev.\ Lett.\  {\bf 112} (2014) 203002
  doi:10.1103/PhysRevLett.112.203002\\{}
  [arXiv:1406.4979 [physics.atom-ph]].
  
  \bibitem{Uzan:2010}
  J.~P.~Uzan,
  ``Varying constants, gravitation and cosmology'',\\
  Living Rev.\ Rel.\  {\bf 14} (2011) 2
  doi:10.12942/lrr-2011-2
  [arXiv:1009.5514 [astro-ph.CO]].
  
  \bibitem{Damour:2001}
  T.~Damour,
  ``Questioning the equivalence principle'',\\
  Compt.\ Rend.\ Acad.\ Sci.\ Ser.\ IV Phys.\ Astrophys.\  {\bf 2} (2001) no.9,  1249\\
  doi:10.1016/S1296-2147(01)01272-0
  [gr-qc/0109063].
  
  \bibitem{Damour:1995}
  T.~Damour and D.~Vokrouhlicky,
  ``The equivalence principle and the moon'',\\
  Phys.\ Rev.\ D {\bf 53} (1996) 4177
  doi:10.1103/PhysRevD.53.4177
  [gr-qc/9507016].
  
  \bibitem{Nordtvedt:1971}
  K.~Nordtvedt,
  ``Equivalence principle for massive bodies. 4. Planetary bodies and modified Eotvos-type experiments'',
  Phys.\ Rev.\ D {\bf 3} (1971) 1683.\\
  doi:10.1103/PhysRevD.3.1683
  
  
  \bibitem{Seveso}
  Luigi Seveso and Matteo Paris, \\
  ``Can quantum probes satisfy the weak equivalence principle?'',\\
  Annals of Physics {\bf380} (2017) 213--223. 
  
  \bibitem{Viola}
  Lorenza Viola and Roberto Onofrio,\\
  ``Testing the equivalence principle through freely falling quantum objects'',\\
  Physical Review D {\bf55} (1997) 455-562. 
  
  \bibitem{Onofrio}
  Roberto Onofrio and Lorenza Viola,\\
  ``Gravitation at the mesoscopic scale'',\\
  Modern Physics Letters {\bf A12} (1997) 1411--1417.
 

\bibitem{schrodinger-newton}
 I.~M.~Moroz, R.~Penrose and P.~Tod,\\
  ``Spherically symmetric solutions of the Schrodinger--Newton equations'',\\
  Class.\ Quant.\ Grav.\  {\bf 15} (1998) 2733.
  doi:10.1088/0264-9381/15/9/019
  
  
   \bibitem{emperor}
 Roger Penrose, \emph{The Emperor's New Mind}, (Oxford, 1989).

  \bibitem{shadows}
   Roger Penrose, \emph{Shadows of the Mind}, (Oxford, 1994).
   
 \bibitem{reality}
  Roger Penrose, \emph{The Road to Reality}, (Alfred Knopf, 2004).

  
  \bibitem{Diosi:1984}
  L.~Di\'osi,
  ``Gravitation and quantum-mechanical localization of macro-objects'',\\
  Phys.\ Lett.\ A {\bf 105} (1984) 199
  doi:10.1016/0375-9601(84)90397-9\\{}
  [arXiv:1412.0201 [quant-ph]].
  
  \bibitem{Diosi:1987}
  L.~Di\'osi,
  ``A universal master equation for the gravitational violation of quantum mechanics'',
  Phys.\ Lett.\ A {\bf 120} (1987) 377.
  doi:10.1016/0375-9601(87)90681-5
  
  \bibitem{Diosi:2014}
  L.~Di\'osi,
  ``Newton force from wave function collapse: speculation and test'',\\
  J.\ Phys.\ Conf.\ Ser.\  {\bf 504} (2014) 012020
  doi:10.1088/1742-6596/504/1/012020\\{}
  [arXiv:1312.6404 [quant-ph]].
  
 \bibitem{GRW} 
  G. C. Ghirardi, A. Rimini, and T. Weber. \\
  ``Unified dynamics for microscopic and macroscopic systems'', \\
  Phys. Rev. D. {\bf34} (1986) 470--491.

\bibitem{bohm1}
D.~Bohm,\\
  ``A Suggested interpretation of the quantum theory in terms of hidden variables. 1.'',\\
  Phys.\ Rev.\  {\bf 85} (1952) 166.
  doi:10.1103/PhysRev.85.166
  
 \bibitem{bohm2} 
  D.~Bohm,\\
  ``A Suggested interpretation of the quantum theory in terms of hidden variables. 2.'',\\
  Phys.\ Rev.\  {\bf 85} (1952) 180.
  doi:10.1103/PhysRev.85.180
  
\bibitem{Dent:2008}
  T.~Dent,
  ``E\"otv\"os bounds on couplings of fundamental parameters to gravity'',\\
  Phys.\ Rev.\ Lett.\  {\bf 101} (2008) 041102
  doi:10.1103/PhysRevLett.101.041102\\{}
  [arXiv:0805.0318 [hep-ph]].

\bibitem{Haugan:1976}
  M.~P.~Haugan and C.~M.~Will,
  ``Weak Interactions and E\"otv\"os Experiments'',\\
  Phys.\ Rev.\ Lett.\  {\bf 37} (1976) 1.
  doi:10.1103/PhysRevLett.37.1
  
  \bibitem{Haugan:1977}
  M.~P.~Haugan and C.~M.~Will,
  ``Principles of equivalence, E\"otv\"os experiments, and gravitational redshift experiments: The free fall of electromagnetic systems to post-Coulombian order'',\\
  Phys.\ Rev.\ D {\bf 15} (1977) 2711.
  doi:10.1103/PhysRevD.15.2711
  
  \bibitem{Kraiselburd:2011}
  L.~Kraiselburd and H.~Vucetich,\\
  ``Constraining the fundamental interactions and couplings with E\"otv\"os experiments'',\\
  Phys.\ Lett.\ B {\bf 718} (2012) 21
  doi:10.1016/j.physletb.2012.10.016\\{}
  [arXiv:1110.3527 [gr-qc]].
  
  \bibitem{Schmole:2016}
  J.~Schm\"ole, M.~Dragosits, H.~Hepach and M.~Aspelmeyer,
  ``A micromechanical proof-of-principle experiment for measuring the gravitational force of milligram masses'',\\
  Class.\ Quant.\ Grav.\  {\bf 33} (2016)  125031
  doi:10.1088/0264-9381/33/12/125031\\{}
  [arXiv:1602.07539 [physics.ins-det]].


\bibitem{COW1}
 R.~Colella, A.~W.~Overhauser and S.~A.~Werner,\\
  ``Observation of gravitationally induced quantum interference'',\\
  Phys.\ Rev.\ Lett.\  {\bf 34} (1975) 1472.
  doi:10.1103/PhysRevLett.34.1472
  
  \bibitem{COW2}
  J.-L.~Staudenmann, S.~A.~Werner, R.~Colella and A.~W.~Overhauser,\\
  ``Gravity and inertia in quantum mechanics'',\\
  Phys.\ Rev.\ A {\bf 21} (1980) no.5,  1419.
  doi:10.1103/PhysRevA.21.1419

\bibitem{Rosi-et-al}
G. Rosi,	F. Sorrentino,	L. Cacciapuoti,	M. Prevedelli	\& G. M. Tino\\
``Precision measurement of the Newtonian gravitational constant using cold atoms'',\\
Nature {\bf 510} (26 June 2014) 518--521 doi:10.1038/nature13433

\bibitem{bounce1}
  V.~V.~Nesvizhevsky {\it et al.},\\
  ``Quantum states of neutrons in the Earth's gravitational field'',\\
  Nature {\bf 415} (2002) 297.
  doi:10.1038/415297a
  
 \bibitem{bounce2}
  V.~V.~Nesvizhevsky {\it et al.},\\
  ``Measurement of quantum states of neutrons in the Earth's gravitational field'',\\
  Phys.\ Rev.\ D {\bf 67} (2003) 102002
  doi:10.1103/PhysRevD.67.102002
  [hep-ph/0306198].
  
  \bibitem{bounce3}
  V.~V.~Nesvizhevsky {\it et al.},\\
 \leftline{ ``Reply to: Comment on `measurement of quantum states of neutrons in the earth's gravitational field'',}
  Phys.\ Rev.\ D {\bf 68} (2003) 108702.
  doi:10.1103/PhysRevD.68.108702


\bibitem{page}
Don N. Page, and C. D. Geilker, ``Indirect evidence for quantum gravity'', \\
Phys. Rev. Lett. {\bf47} (1981) 979--982. \; doi:10.1103/PhysRevLett.47.979

\bibitem{Kiefer:2008}
  M.~Albers, C.~Kiefer and M.~Reginatto, ``Measurement analysis and quantum gravity'',\\
  Phys.\ Rev.\ D {\bf 78} (2008) 064051
  doi:10.1103/PhysRevD.78.064051\\{}
  [arXiv:0802.1978 [gr-qc]].





\end{thebibliography}
\end{document}